\documentclass[vecphys]{svmult}

\usepackage{graphicx}        
\usepackage[bottom]{footmisc}

\begin{document}

\title*{Jets and Tori in Proto-Planetary Nebulae: \\
  Observations vs. Theory}
\titlerunning{Jets and Tori}
\author{P. J. Huggins}
\institute{Physics Department, New York University, New York, NY 10003,
  USA \\
\texttt{patrick.huggins@nyu.edu}}
%
%
\maketitle

\begin{abstract}
We report on a study of the time sequence for the appearance of
high-velocity jets and equatorial tori in the transition of stars from
the asymptotic giant branch to the planetary nebulae phase. Jets and
tori are prominent features of this evolution, but their origins are
uncertain. Using the kinematics of molecular tori and molecular or
optical jets, we determine the ejection histories for a sample of
well-observed cases.  We find that jets and tori develop nearly
simultaneously. We also find evidence that jets appear slightly later
than tori, with a typical jet-lag of a few hundred years. The
reconstructed time-lines of this sequence provide good evidence that
jets and tori are physically related, and they set new constraints on
jet formation scenarios.  Some scenarios are ruled out or rendered
implausible, and others are challenged at a quantitative level.

\end{abstract}

\section{Introduction}
\label{sec:1}

Jets and equatorial tori are among the most prominent morphological
features of proto-planetary nebulae (proto-PNe). They are also
relatively common, but their origins are uncertain. In this paper we
attempt to make connections between the observed characteristics of
the jets and tori, and various theoretical ideas on how they might
form.
 
The paper is divided into two parts. In the first part we focus on
observations, and address the question whether jets and tori are
related in some way.  We examine this in the time domain, by asking if
there is a consistent pattern for the ejections, and what the
relevant time scales are. We conclude that jets and tori are related:
they develop nearly simultaneously, and we find evidence for a
torus-jet sequence.  In the second part of the paper, we use this
torus-jet relation to evaluate different formation scenarios. It
favors and constrains certain scenarios, and rules out others.

\section{Observations}
\label{sec:2}
 
The launching of high-velocity jets and the ejection of equatorial
tori are among the most traumatic events in the lives of stars in the
transition from the asymptotic giant branch (AGB) to the PN phase.
Figure~1 illustrates the relatively sudden change in geometry from the
regular AGB mass-loss.  The left hand panel shows a simplified
picture, and the right hand panel shows a real example, AFGL~618,
imaged with the HST. The jets of AFGL~618 show multiple components
(the cause is not known), but they have well defined tips, and there
is a dense torus, seen in absorption around the equator, which was
formed by the last major mass-loss event from the star.  It might come
as no surprise that such dramatic mass ejections are somehow related,
but the question has not previously been addressed in any detail.

\begin{figure}
\centering
\hspace{-1.5cm}
\includegraphics[height=5.0cm]{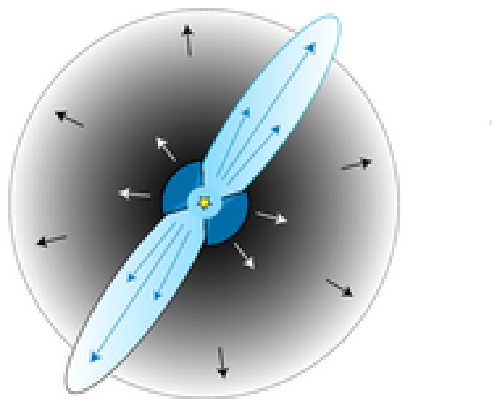}
\includegraphics[height=5.0cm]{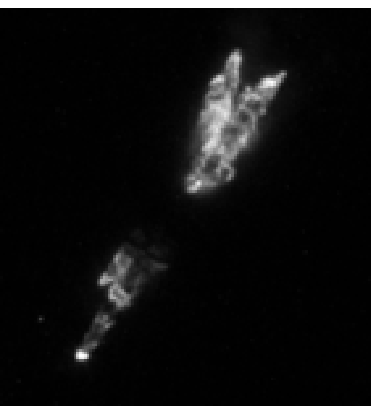}
\caption{Jets and tori: ({\bf left}) simplified picture; ({\bf right})
 the proto-PN AFGL~618, imaged by the Hubble Space Telescope, adapted from
 \cite{tra02} }
\label{fig:1}       
\end{figure}

\subsection{Kinematic Ages} 
We explore the relation between jets and tori in the time domain by
estimating when each component was ejected. We can determine the
kinematic ages of the jets from their velocities and radial
extent. This can be done using imaging and spectroscopy of optical
emission lines or molecular lines if there are molecules in the
jets. The estimates require information on the tilt of the system
(because only projected quantities are measured), and in well studied
cases this can be determined from geometrical or kinematic
considerations. Observations of optical proper motions of the tips of
the jets can also yield kinematic ages, independent of the tilt.
 
For the equatorial outflows, we emphasize in this study objects with
massive molecular tori whose expansion velocity and radial extent have
been observed in millimeter CO lines. This is an important
consideration: tori can of course be studied in optical emission lines
in fully developed PNe, but by then the kinematics of the gas will
have been strongly affected by the onset of ionization, and will no
longer be a reliable indicator of the time since ejection.  We note
that the ejection of the tori in some cases may be more spherically
symmetric than suggested in out schematic figure, but pierced by the
jets: this does not appreciably affect our discussion. We also note
that in the classic paper \cite{sok00}, Soker \& Rappaport argued that
tori are not distinct ejections, but are simply formed by jets
snow-plowing the AGB wind towards the equatorial plane; however the
high masses and high mass loss rates typically found, e.g.,
\cite{hug04}, argue for something different.
 
There are nine objects with observations suitable for comparing the
kinematic ages of jets and tori. They range from AGB stars to young
PNe: $\pi^1$~Gru, AFGL~618, V~Hya, He~3-1475, M~2-56, M~1-92, M~2-9,
M~1-16, KjPn~8. We have omitted a few well-observed cases where the
geometry is complex, e.g., AFGL~2688 \cite{cox00}. The data for the
sample come from the work of several research groups and the details
are given in \cite{hug07} (see also J.~Alcolea, this volume).  For this
sample, the median expansion velocity of the tori is 10\,km\,s$^{-1}$
(comparable to or less than the wind velocity near the end of the
AGB), and the median jet velocity is 160\,km\,s$^{-1}$. The kinematic
ages range from 50 to 5,000\,yr, and they roughly correlate with the
evolution of each object suggested by the morphology of the nebula
and/or the temperature of the central star; this is consistent with
all ejections occurring near the end of the AGB.

\subsection{Jet-Lag}
The kinematic ages of the jets are plotted against those of the tori
in the left hand panel of Fig.~2. In spite of the relatively crude
measures involved (especially of the tori which are not highly
resolved) and the uncertainties in the orientations (see \cite{hug07}
for details), the ages are seen to be correlated. The jets and tori
occur nearly simultaneously, strongly suggesting they are physically
related.
 
Further inspection of the figure shows that the kinematic ages of the
jets are slightly less than those of the tori. There is good evidence
to suggest that the kinematic ages are quite reasonable estimates of
the true travel times: for the tori because of their large inertia and
low velocity, and for the jets because of the Hubble-like ballistic
flows seen in well studied cases, e.g., \cite{alc01,cor04,uet06}.  On
this basis, the most reasonable interpretation of the figure is that
the jets typically develop after the tori, with a delay or
jet-lag. From the age measurements for the ensemble of objects, the
typical jet-lag is 300\,yr.

Although there is good evidence for approximate, unimpeded ballistic
motion for the jets, the dynamics is not understood in detail, so
there could be slight variations. The difference between the true age
and the kinematic age can be seen by writing the former as $t = \int
dr/v(r)$ (where $v$ is the velocity as a function of distance $r$),
whereas the latter is $r/v$ at the current epoch. To the extent that
the jets are decelerated by interaction with the circumstellar gas,
their true ages will therefore be shorter than the kinematic ages, and
the jet-lag will be longer than we estimate.  Acceleration of the jets
is unlikely to dominate the kinematics (see \cite{hug07} for details),
but the jet-lag could be somewhat shorter. In any event, our
measurements refer to the tips of the jets and characteristic or
average radii of the tori, so our finding that the jets are almost
simultaneous or delayed relative to the tori is likely to be
robust. This sequence is also consistent with the kinematics and
morphology seen in individual cases, e.g., \cite{hug04}, where
material of the torus is entrained along the sides of the jets.

\begin{figure}
\centering
\includegraphics[height=5.0cm]{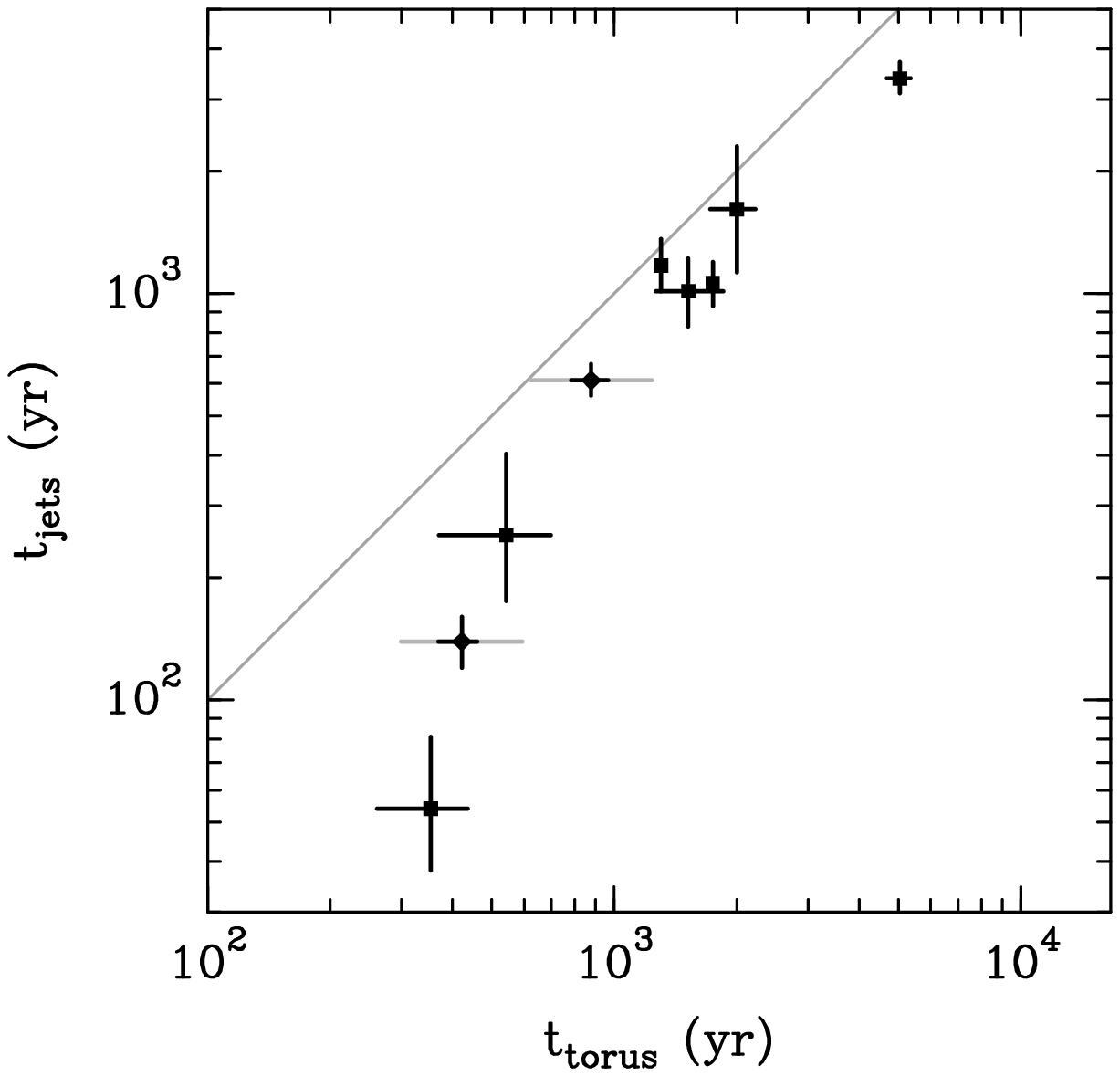}
\hspace{0.5cm}
\includegraphics[height=5.0cm]{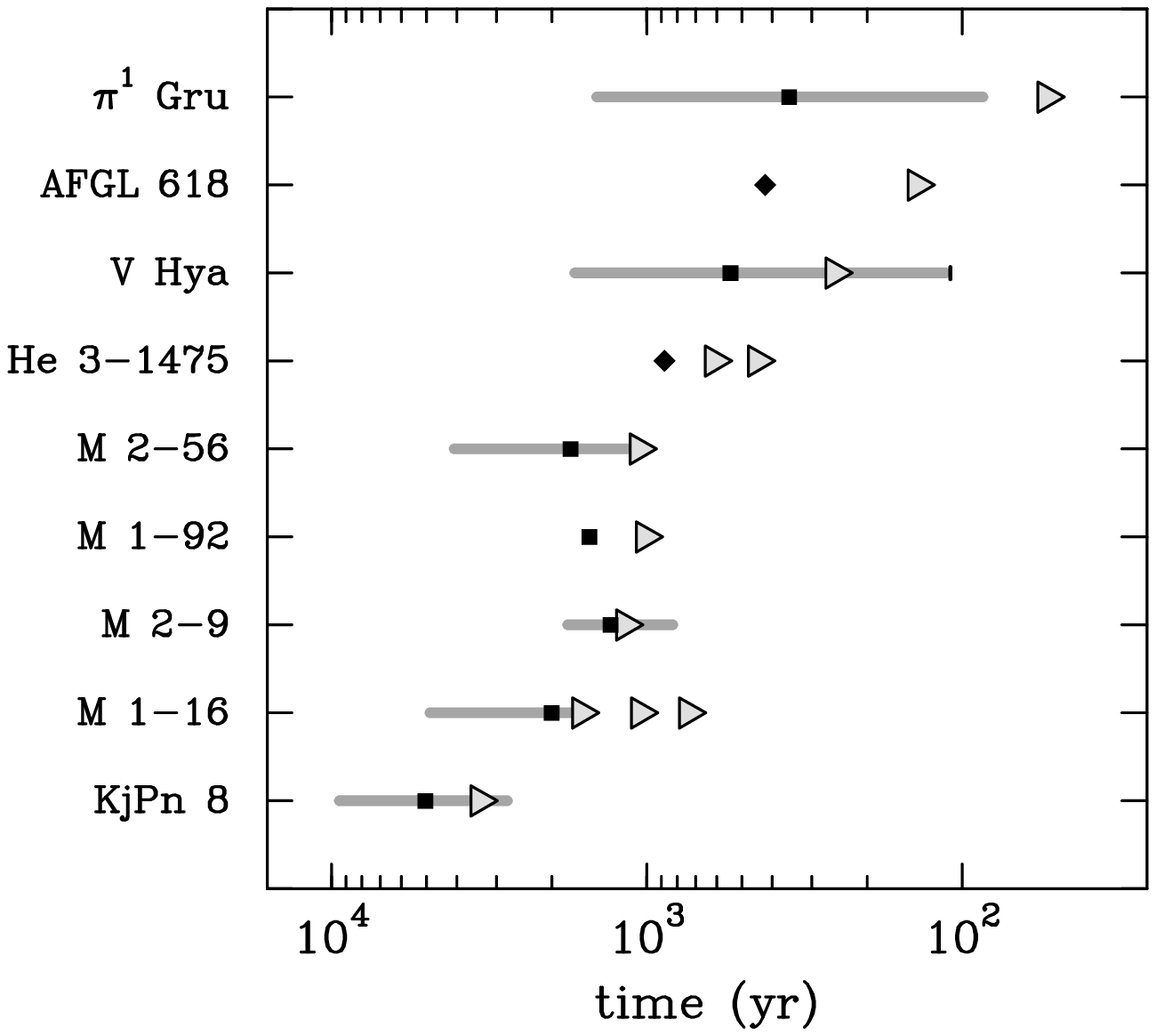}
\caption{The timing of jets and tori: ({\bf left}) kinematic ages of jets
  vs.\ tori (error bars indicate uncertainties in inclinations, proper
  motions, etc., and the continuous line shows where the ages are
  equal); ({\bf right}) reconstructed time lines, with time axis in years
  ago from now (horizontal lines denote the duration of torus
  ejections, and arrowheads denote jet ejections)}
\label{fig:2}       
\end{figure}

\subsection{Torus-Jet Sequence}
 
Using our timing estimates we can reconstruct the evolution of the
outflows, and this is shown in the right hand panel of Fig.~2.  The
horizontal axis is time in the past, measured from the current epoch.
The filled points show when the tori were ejected, and the horizontal
lines show estimates for the duration of the ejection, based on the
radial widths of the tori and the expansion velocity.  These would be
overestimates if there is an appreciable internal dispersion in the
gas. The arrowheads show when the jets were launched.  In two cases
there are data for multiple jets, and the interval between them is
seen to be comparable with the jet-lag.
 
In spite of the considerable heterogeneity of the data used to make
Fig.~1, the overall picture for the ensemble of objects is remarkably
consistent. Tori are ejected on relatively short time scales, and jets
are launched at the same time or shortly after.

\section{Implications}
 
We now turn to the implications of these results for ideas on jet
formation. There are several theoretical ingredients, MHD launching
and collimation, disks of various kinds (primary, secondary,
circum-binary), the effects of a common envelope phase, etc., and they
come together in four types of scenario which we discuss in turn. 

\subsection{Scenarios}
Each scenario has fairly specific predictions on the relations between
the jets and tori which we can test using the findings from the
observations described in the previous section.
 
\paragraph{Single Stars} 
The first scenario involves magnetic winds from single stars. Current
models with a strong enough magnetic field can produce jets
\cite{gar05} although there are likely problems with this idea
\cite{sok06}. In the context of our present results, it is unclear if
single star models can produce sudden jets, and those available do not
produce co-ordinated tori. The equatorial density enhancements in the
simulations by \cite{gar05} are input separately, and do not
constitute a discrete ejection. In view of the close connection
between jets and tori found here, the current single star models do
not give a good description of the observations.

\paragraph{Disks around Binary Companions} 
The second scenario involves the launching of jets from accretion disks around
binary companions, e.g., \cite{mor87,sok00}, similar to the jet
launching picture
in young stellar objects. In this case the disk is fed by the
mass-loss from the AGB primary. This scenario has several attractive
features. First, it provides a natural mechanism for enhanced mass-loss in the
equatorial plane. Second, it provides a causal and sequential relation between
the torus and jets, because the enhanced mass-loss of the torus feeds
the accretion disk that in turn fuels or triggers the jets. It even provides a
natural explanation for jet-lag. 

Jets will not immediately respond to enhanced mass-loss and enhanced
accretion, but will lag by the time it takes the matter to spiral into
the companion. This accretion time is the viscous timescale of the
disk, which is given by the following expression in the usual
$\alpha$-prescription:

\begin{equation}
t_{\nu} = 160\,\mathrm{yr} \, \left(\frac{\alpha}{0.1}\right)^{-1}
\left(\frac{R}{1\, \mathrm{AU}}\right)^{3/2} \left(
\frac{M_2}{1\,M_\odot} \right)^{-1/2} \left( \frac{H/R}{0.1}
\right)^{-2}  ,
\end{equation}
where $\alpha$ is the viscosity parameter, $M_2$ is the mass of the
companion, and $H$ and $R$ are the scale height and radius of the
disk, respectively. For reasonable values of the parameters (given by
the scaling values in the equation), the viscous timescale is a few
hundred years, comparable with the typical jet-lag found from the
observations. Thus the jet-lag could well be the signature for the
presence of an accretion disk (which is not otherwise detected) and
can provide quantitative information on its properties.

In spite of the success of this scenario, one feature that is not
generally explained is the onset of a discrete torus. It may be
connected with spin-up by the secondary, and/or a critical envelope
mass of the primary. This point warrants further study.
 
\paragraph{Spun-up Envelopes}  
The third type of scenario involves the effects of spin-up of the
envelopes, e.g., \cite{nor06,mat06}, especially during a common envelope
phase which might naturally lead to ejection in the equatorial plane. One
version involves common-envelope ejection to form a torus, while the
spin-up of the envelope generates jets.  A second version involves
the build-up of the stellar magnetic field by rotational shear until
it explodes in both the polar and equatorial directions. This case is
interesting in view of the possible observable characteristics of the
magnetic field in the tori, e.g., \cite{hug05,sab07}.

For both of these scenarios, the time scale for the sequence is likely
to be fairly rapid, so it is unclear if either of them is consistent
with jet-lag of a few hundred years. Similarly, it is unclear whether
common-envelope ejection or a magnetic explosion can give rise to the
low velocity tori: in all observed cases the bulk of the mass comes
off with a velocity comparable to or lower than the wind velocity on
the AGB. Realistic simulations of the ejection process in a common
envelope need to be developed.

A hybrid model, in which a binary companion first accretes matter and blows
jets, and is then engulfed by the primary to form a common envelope
phase that ejects a torus is ruled out: it gives the wrong torus-jet
sequence.
 
\paragraph{Disks around the Primary Core}  
The fourth type of scenario involves the production of jets from an
accretion disk around the primary core or core-remnant
\cite{sok94,sok96,rey99,nor06}. One version involves a low mass
object that is engulfed by the primary and is tidally disrupted to form a
disk-jet system around the core, to be later followed by the
normal evolution of the star and the ejection of the nebula. This
scenario can be ruled out because the jets and torus are uncoordinated
(contrary to observations) and they occur in the wrong sequence.

A second possibility is for a companion to eject a torus in a common
envelope phase, followed by tidal disruption forming a
disk-jet system around the core. This generates the correct torus-jet
sequence, but is likely restricted to a narrow range of companion mass,
and the time scales are uncertain.

A third possibility is one in which a torus is formed by common
envelope ejection, and then the secondary undergoes Roche lobe
overflow, feeding a disk-jet system around the remnant core. This
gives the correct torus-jet sequence, but the time scale on which the
disk is formed is likely too long, and it can probably be ruled out.

\subsection{Summary of Scenarios}
As seen by the above discussion, the torus-jet connection provides
important constraints on possible scenarios.  Table~1 provides a
summary, with a rating for each of the cases considered.  Half of them
are ruled out or made implausible, and the remainder need more
realistic simulation for comparison with the observed constraints.

All the plausible scenarios involve a stellar or sub-stellar mass
companion. Given the ubiquity of jets, this would imply interactions
on the AGB should be a common phenomenon, and there is growing
evidence that this might be the case, e.g., \cite{mau06} (see also
R.~Sahai, this volume).

 \begin{table}
\centering
\caption{Jet-Torus Scenarios}
\label{tab:1}       
\begin{tabular}{lcl}
\hline\noalign{\smallskip}
Scenario & \ \ Rating\ \ \  & Comments \\
\noalign{\smallskip}\hline\noalign{\smallskip}
mag.\ wind from single star            & $-$  & jets \emph{and} torus?    \\
primary mass loss + companion acc.\ disk  & \ + \ &  discrete torus
ejection?   \\
companion acc.\ disk + CE ejection & $-$ &  wrong sequence  \\
CE ejection + mag.\ polar wind    & \ + \ &  jet-lag?   \\
(CE) mag.\ polar \& equatorial explosion   &  \ + \  &  jet-lag?   \\
(CE) primary acc.\ disk + late PN   & $-$ &  wrong
sequence   \\
CE ejection + primary acc. \ disk  &   \ + \ &  jet-lag?   \\
CE ejection + RLOF    & $-$  &
time scale too long?    \\
\noalign{\smallskip}\hline
\end{tabular}
\end{table}

\section{Conclusions}
 
The results of this study show that the launching of jets and the
ejection of equatorial tori are related. They are nearly simultaneous,
and there is evidence for a torus-jet sequence with a typical delay
time or jet-lag of a few hundred years.

The near simultaneity of the outflows, the torus-jet sequence, and
time scales for their development set interesting constraints on
scenarios for jet and torus formation.  The observations already rule
out or make implausible several proposed scenarios, and for others
they pose well defined, quantitative questions that need to be addresses
by realistic simulations.

\subparagraph{Acknowledgments.} This work was supported in part by NSF
grant AST 03-07277.

%
%
%



\end{document}